\documentclass[showpacs,preprintnumbers,amsmath,amssymb]{revtex4}
                                                                                                              
\usepackage{graphicx}
\usepackage{subfigure}
\begin{document}
                                                                                
\centerline{\bf  Queue-length Variations In A Two-Restaurant Problem}
\vskip 1 cm                                                                                
\centerline{Anindya S. Chakrabarti,$^{1}$}

\centerline{Bikas K. Chakrabarti$^{1,2}$}

\centerline{\small $^{1}$Economic Research Unit, Indian Statistical Institute,  }

\centerline{\small  203 B.T.Road,Kolkata-700 018,India}

\centerline{\small $^{2}$Center for Applied Mathematics and Computational Science, }

\centerline{\small Saha Institute of Nuclear Physics, }

\centerline{\small 1/AF Bidhannagar, Kolkata-700 064, India.}

\vskip 1 cm

\noindent {\bf Abstract:}
{\small This paper attempts to find out numerically the distribution of the queue-length ratio in the context of
 a model  
of preferential attachment. Here we consider two restaurants only and a large number of customers 
(agents) who come to these restaurants. Each day the same number of agents sequentially arrives and decides 
which restaurant to enter. If all the 
agents literally follow the crowd then there is no difference between this model and the famous `P\'olya's Urn' model.
 But as agents alter their strategies different kind of dynamics of the model is seen. It is seen from numerical 
results that the existence of a distribution of the fixed points is quite robust and it is also
seen that in some cases the variations in the ratio of the queue-lengths follow a power-law.
}

\section{Introduction}

\noindent There are many social or financial contexts where individual rationality working 
alongwith the force of 
imitation gives rise to a situation where all people are `worse-off' in the Paretian 
sense than they were 
before. This apparently paradoxical process of `mimetic rationality' has been captured in the
works of Banerjee [1] and Orl\'ean [2](see also [3], [4]). Economists use the term `informational cascade' to 
depict such a flow of information. `P\'olya's Urn' model (see e.g. [5], [9]) plays an important role here.
Following P\'olya, we present a very simplified version of a model of preferential attachment. This is an 
N-agent game, where N is a sufficiently large number. There are two restaurants. Each agent without communicating
with others sequentially decides which restaurant to enter. None of these agents have any inclination 
towards any particular 
restaurant. Their choices are guided by their peers' choices only. Next, we see that if they start to depend 
on the history of the game
or if they have any outside information then the dynamics of the game is seen to get 
altered and the distribution of the fixed 
points also becomes altered. It has been claimed that the ratio of the queue-lengths
measured from hospital-waiting lists follow a power-law decay [6], implying the presence of complexity.
But complexity, though a sufficient condition to exhibit power law, is not a necessary one as has been argued
in [7].  
Here we show that one would get similar data-sets exhibiting power-laws just by recasting 
the urn model in the social context.

\section{The Problem}

\noindent We consider two restaurants which are identical in all respects. Initially these two restaurants
are occupied by one agent each. At every time-step one agent arrives and decides which restaurant to enter.
Suppose there are $N_A$ number of customers in restaurant A and $N_B$ number of customers in
 restaurant B. We assume that 
the probability that the next agent assigns to restaurant A is {$N_A$}/{($N_A$+$N_B$)} and the 
rest is assigned to restaurant B. Naturally there will be some fluctuations in the occupation 
density of A and B. But eventually these
fluctuations die out and the system reaches an equilibrium where there exists a fixed point ($P_A$) 
indicating the share of the agent-population selecting restaurant A (i.e. the equilibrium occupation 
density of
restaurant A). Clearly for restaurant B the
fixed point is $P_B=1-P_A$.

\indent The game considered above is assumed to end after reaching the equilibrium. If the game is played
independently for a sufficiently large number of time then we see that there are a large number of
 fixed points for this game.
 It is in fact a welknown result that there exists an infinite number of fixed  points ($P_A$) for 
this game distributed 
uniformly over $[0,1]$(P\'olya's Urn Model) (see ref. [3], [5]). 
Suppose each day $N$ number of agents arrive and the number of agents gone 
to restaurants A and B at the end of the day, is assumed to be the queue-lengths($Q_A$ and $Q_B$). This game is 
played for $D$ days and the queue-lengths on the i-th day are denoted as ($Q_{Ai}$ and $Q_{Bi}$). 
 Next, we consider the cases where there is a general perception that a particular restaurant among 
A and B is better 
than the other or where agents 
arbitrarily change their opinion about the best choice (the reason might be that there is no clear-cut winner) 
and also the case where agents are influenced by history. 

First, we intend to find out the distribution of the queue-length in different variations of this two-restaurant 
problem. Ultimately we are concernd with the distribution of the ratio of the queue-lengths i.e. the distribtion of 
$L_A=(Q_{Ai}/Q_{Aj})$ and $L_B=(Q_{Bi}/Q_{Bj})$ where $j = i+1$. Clearly the ratio of the queue-lengths
defined thus is the ratio of the fixed points. In the numerical simulations below we plot the distribution
of $P_A$ i.e. the distribution of the occupation densities 
of restaurant A after the system has reached equilibrium. 
Clearly $P_B=1-P_A$ in all cases considered below. The queue-lengths can also be expressed as
$Q_A=N.P_A$ and $Q_B=N.P_B$.

\section{\it Stochastic Strategies For Agents }
\indent{\bf Case (i)} {\it Agents avoid crowd} : Suppose after the game has been
played by a certain number of agents, $N_A$ number of agents have gone to restaurant A 
and $N_B$ number of agents have gone to B. The next 
agent assigns probability $N_A/(N_A+N_B)$ to restaurant B and the rest to A. 
If agents avoid the crowd this way then eventually the dynamics is 
absorbed in a stable attracting
 fixed point $P_A=0.5$ after sufficient number of iterations. 
This result can readily be generalized to any number of restaurants.
\medskip  

\indent{\bf Case (ii)} {\it Agents choose randomly} : If agents choose randomly then eventually the dynamics 
is absorbed in the
 fixed point $P_A=0.5$. This result also can readily be generalized to any number 
number of restaurants.
\medskip 

\indent{\bf Case (iii)} {\it Agents follow the crowd only} : Suppose there are $N_A$ number of customers in 
restaurant A and $N_B$ number 
of customers in restaurant B. We also assume that the probability that the next agent assigns to restaurant A 
is 

\centerline{$P_A=N^\epsilon_A/(N^\epsilon_A+N^\epsilon_B)$  \indent\indent \indent \indent \indent \indent \indent\indent \indent     ...(1)}

and the rest is assigned to restaurant B. $\epsilon$ is a choice parameter.
\medskip

\noindent The numerical results observed are :

\indent (a) For $(\epsilon<1)$ : This leads to a situation where all hotels share the agents equally. 
The fixed point is uniquely determined at $P_A=0.5$.

\indent (b) For $(\epsilon=1)$ : This is the P\'olya model. 
It is a wellknown result that there exists an 
infinite number of fixed  points $P_A$ for this game distributed uniformly over $[0,1]$.

\indent (c) For $(\epsilon>1)$ : One restaurant gradually absorbs the total population. 
Which restaurant will eventually
get all agents will depend upon the choice of the 
first few agents.
\medskip 

See reference [8] for analytical proof.
\medskip

\indent{\bf Case (iv)} {\it Existence of a general preference} : Suppose each agent is not  
influenced by previous choices 
made by their predecessors only but also there exists a general, unanimous perception that either restaurant
 A or restaurant B is better
than the other. For clarity, we assume the general perception is that restaurant A is better 
than restaurant B and we call
$\alpha$ to be the perception coefficient 
where $0 \le \alpha \le 1$. It is to be noted that if $\alpha>1$ then the agents assign a negative 
probability to restaurant B
(since initially there is only a single agent in restaurant B) which is meaningless.
 Hence the probabilities that an agent assigns to restaurant A and B are 

\medskip
\centerline{$P_A=\frac{N_A+\alpha}{N_A+N_B}$  and  $P_B=\frac{N_B-\alpha}{N_A+N_B}$    \indent \indent \indent \indent \indent \indent\indent \indent \indent \indent \indent      ...(2)
}
\medskip
\noindent Numerical results obtained are as follows :

\indent (a) For ($0<\alpha<1$) : Restaurant A attracts more agents as is intuitively 
clear. However, not all agents are  attracted to restaurant A. Fixed points exist in the interval $[0,1]$. 
But this time the spread is not uniform.
 Bulk of the distribution is in the side of A, which grows with $\alpha$. (See fig.1).

\indent (b) For ($\alpha=1$) : Restaurant A attracts all agents from the very begining. This process 
has a single fixed point at $P_A=1$.

\begin{figure}
\begin{center}
\noindent \includegraphics[clip,width= 5cm,angle = 270]
{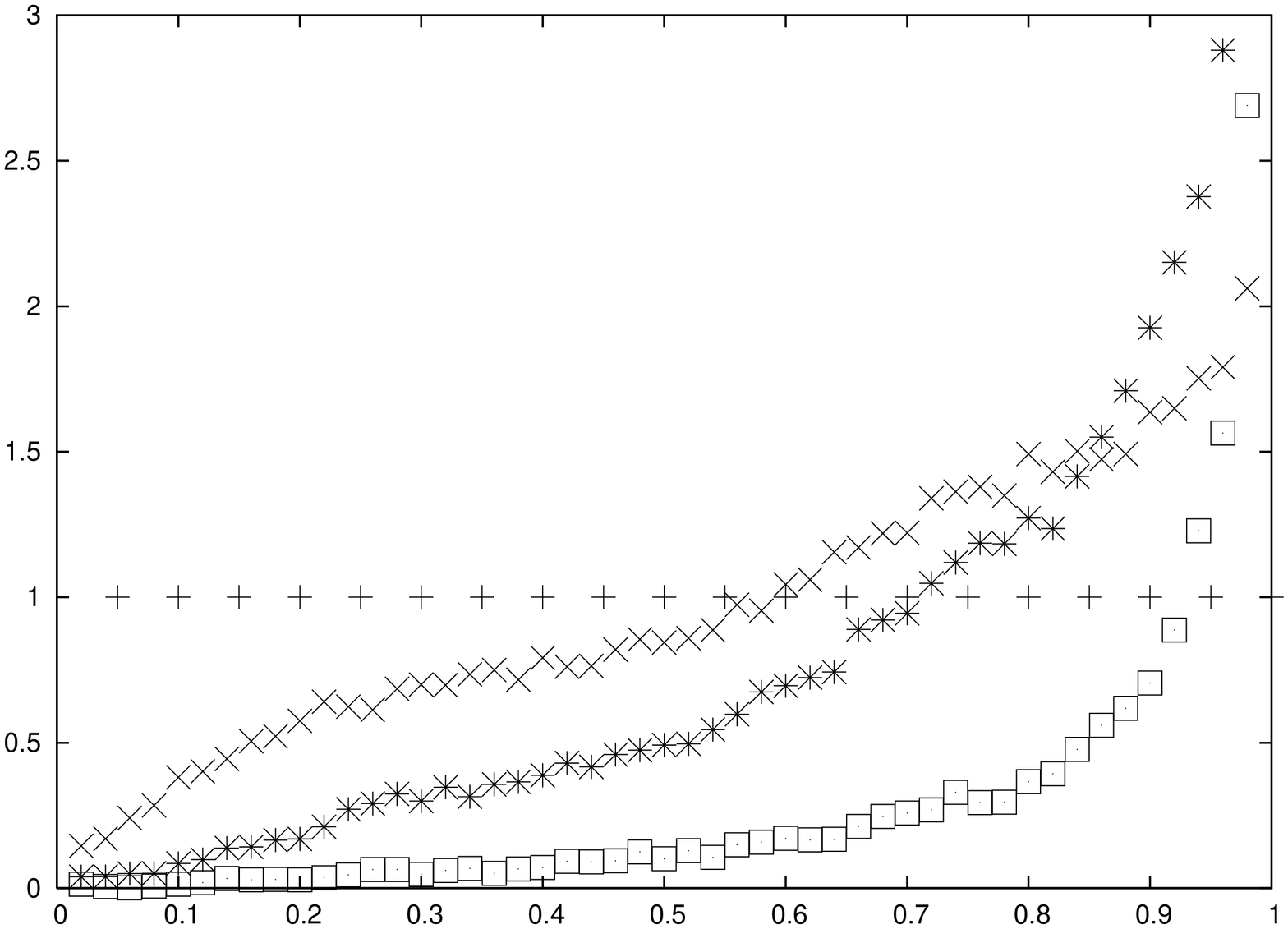}
\caption\protect{\label{fig:fig1}Numerical results of case (iv). $N = 5,000$ and $D = 40,000$ . 
The horizontal points (+) (showing uniform distribution) has 
been drawn for reference. This 
line corresponds to the case where $\alpha=0$ i.e. the P\'olya model.
The distribution of the fixed-points $P_A$ has been plotted for $\alpha=0.3$ ($\times$) 
, $\alpha=0.6$ ($\ *$ ) and $\alpha=0.9$ $ (\square)$ . 
It is seen that as $\alpha$ goes up the bulk of the distribution shifts towards the right 
end showing that restaurant A attracts more agents (but not all).    }

\end{center}
\end{figure}

\medskip 

\indent{\bf Case (v)} {\it Agents have arbitrary preference} : Suppose that each agent individually
prefers either A to B or B to A, though
 broadly speaking they take decisions by following the crowd. To model this we assume the same parameter $\alpha$ 
which this time can be negative and positive with equal probability. In other words, each agent 
considered in this case are equally probable to select A over B or B over A. The probability assigned to 
restaurant A
is   $P_A=\frac{N_A+\alpha}{(N_A+N_B)}$  where $\alpha>0 $ or $\alpha<0$ with the 
same absolute value with equal probability. 
\medskip

\noindent Numerical result obtained is as follows :
\medskip

For small values of $\alpha$ there is almost no distinction between the distribution
of the fixed points and an uniform distribution. But as $\alpha$ goes up the difference becomes more and
more prominent. With increasing value of $\alpha$ the distribution shrinks around the point 0.5.
Ultimately the distribution converges to a $\delta$-function. (See fig.2).
\medskip 

\begin{figure}
\begin{center}
\noindent \includegraphics[clip,width= 5cm,angle = 270]
{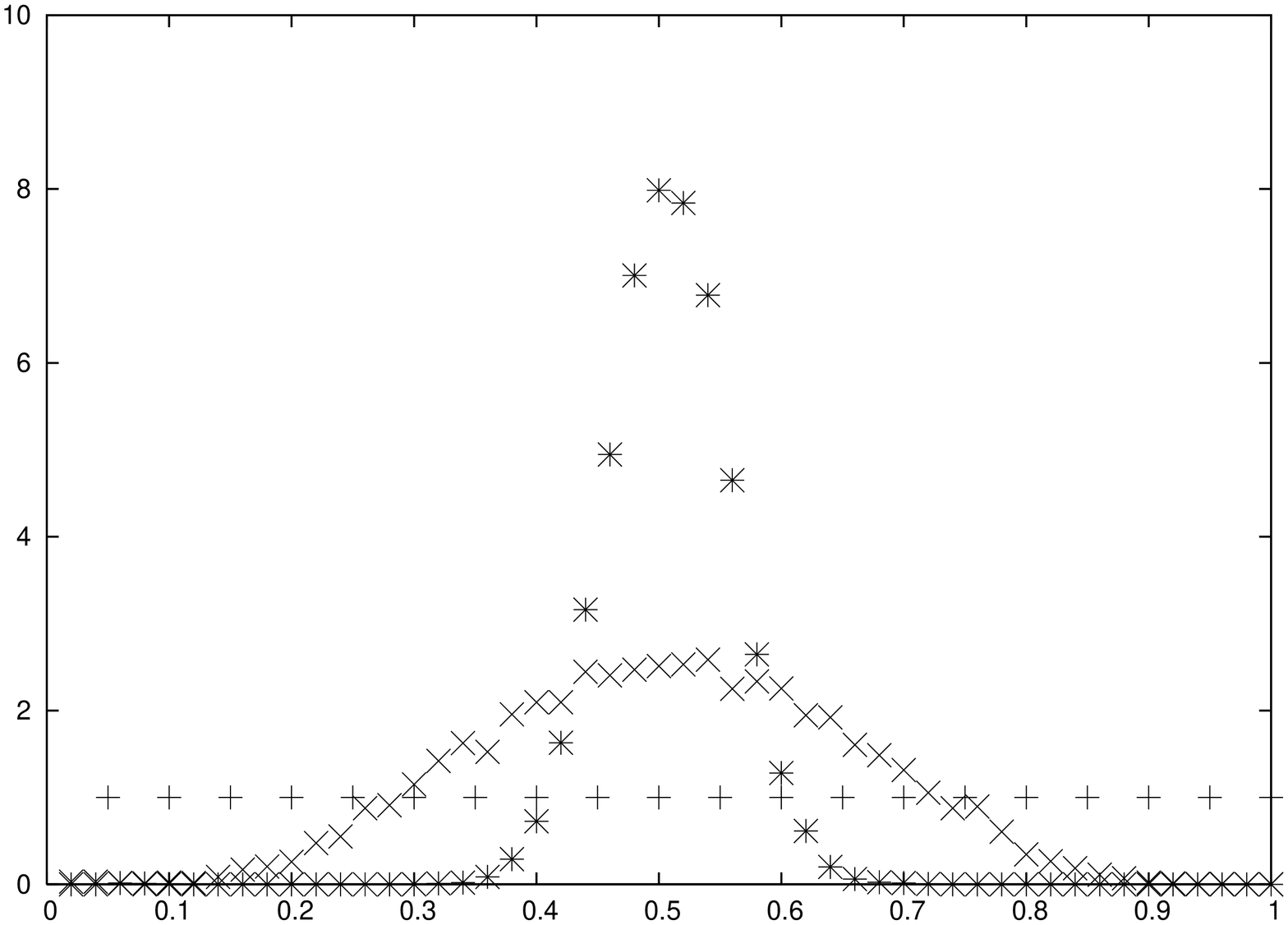}
\caption\protect{\label{fig:fig2}Numerical results for case (v). $N = 5,000$ and 
$D = 10,000$. The distribution of the fixed points 
($P_A$) has been
drawn for $|\alpha|=10$ ($\times$) and  $|\alpha|=100$ ($\ *$ ). As before, the horizontal points (+)
showing the uniform distribution (for $\alpha=0$) has been drawn for reference. 
The contraction of the distribution is apparent with the rises
 in the absolute value of the preference parameter $\alpha$.}
\end{center}
\end{figure}

\indent{\bf case (vi)} {\it Dependence on history} : Here we consider the case where agents not 
only see the current pattern
of the crowd but also give weightage to the restaurants which were more attractive 
in the previous days.
 To model this situation, we assume two parameters $\gamma$ and $\delta$. We call 
$\delta$ to be the
discount factor and $\gamma$ to be the weightage factor (i.e. 
$\gamma$ determines the weightage
given to the present pattern of the crowd to that in the history).

Suppose already the game has been played for $m$ days (i.e. $m$ times). Today a new game has been
started and the fraction of agents gone to A and B are $N_A$ and $N_B$ respectively. In P\'olya 
model the next agent was assigning a probability $P_A= N_A/(N_A+N_B)$ to A and the rest to B.
 Here the agent considers all the fixed points in the previous $m$-rounds of the game. Let's call
those fixed points $P_1, P_2,...    , P_m$ etc. From the agent's point
of view, the most recent history is $P_m$, the second most recent is $P_{m-1}$ etc.
So he calls his most recent history to be $H_1$, the second most $H_2$ and the
last one (i.e. $P_1$) to be $H_m$ etc.
He discounts each period of history by $\delta$.

So history for restaurant A is $H_A = (\delta.H_1+\delta^2.H_2+...  ...+\delta^m.H_m)/Z$
\indent \indent \indent where $Z=(\delta+\delta^2+...  ...+\delta^m)$

\indent \indent\indent\indent\indent\indent
\indent \indent \indent \indent \indent \indent\indent \indent \indent \indent \indent \indent
\indent \indent \indent \indent \indent \indent\indent \indent \indent \indent \indent \indent 
\indent \indent \indent \indent \indent \indent\indent
and $0 \le \delta \le \infty$.
  
Now, $P_A=\gamma.\frac{N_A}{(N_A+N_B)}+(1-\gamma).H_A$  \indent\indent\indent\indent\indent where $0 \le \gamma \le 1$.  \indent \indent \indent \indent \indent \indent\indent \indent \indent \indent \indent \indent
    ...(3)

Clearly, $H_B=1-H_A$ and $P_B=1-P_A$.
It should be clear from the expression of $H$ that as $\delta$ becomes close to zero, only the recent 
history matters. If $\delta=1$,then each round of the game played before is assigned an equal
weightage. As $\delta$ exceeds 1 and goes up, more and more weight is assigned on the begining
of the history. 

\noindent Numerical results obtained are as follows :
\medskip

(a) {\it No dependence on history} ($\gamma=1$) : If $\gamma=1$, 
then only the current status of the game
matters i.e. there is absolutely no dependence on history.
 So it is the P\'olya model basically and the distribution of 
fixed points is uniform over $[0,1]$.

(b) {\it Complete dependence on history} ($\gamma=0$) : In this game again the distribution shrinks to
a $\delta$-function with its peak at 0.5.

(c) {\it Cases in between the above two extremes} : The distribution has both its mean and mode at
0.5 (the distribution is symmetrical about 0.5). The spread of the distribution depends on the 
discount factor $\gamma$ for a given value of $\delta$.

See fig. 3.

\begin{figure}
\begin{center}
\noindent \includegraphics[clip,width= 12cm,angle = 270]
{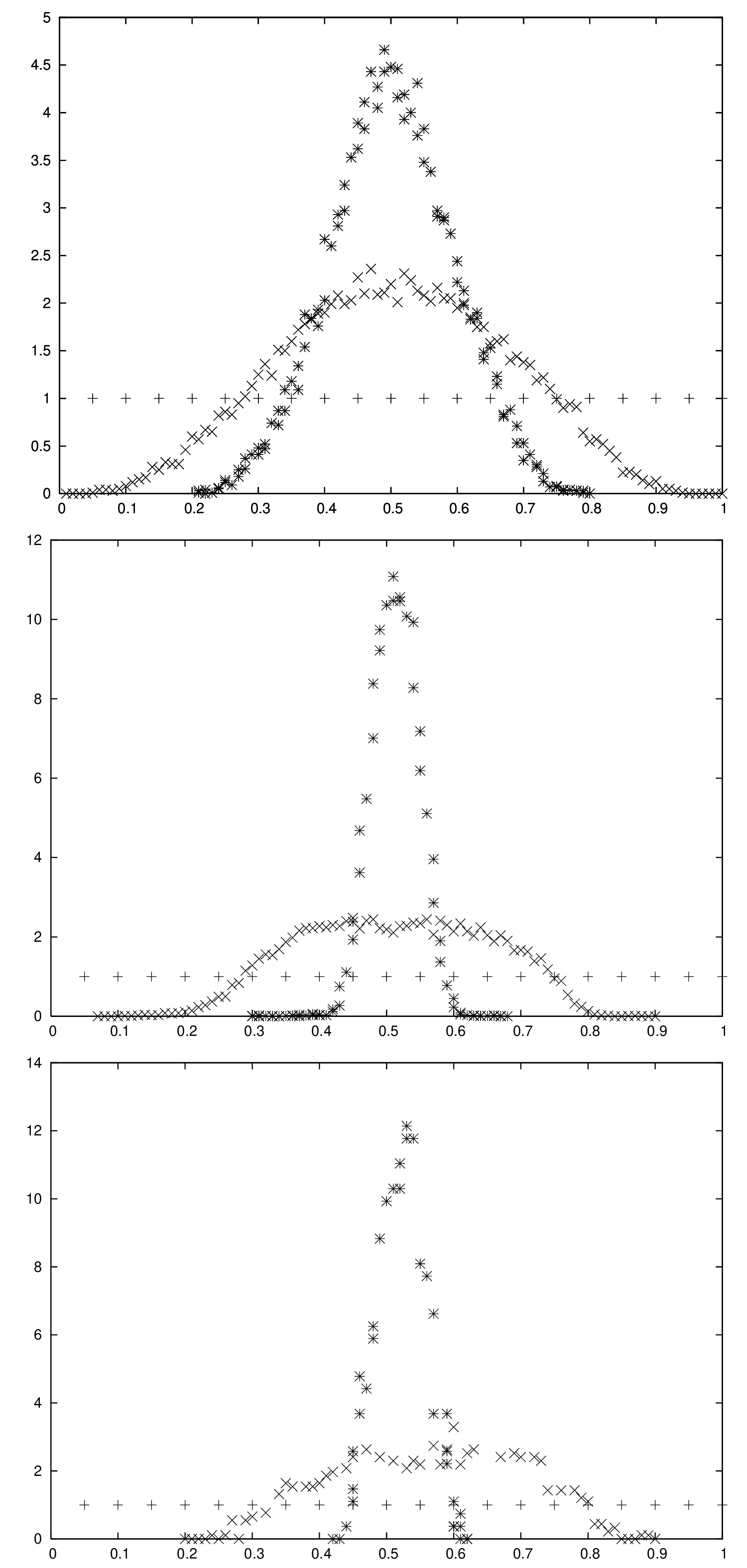}
\caption\protect{\label{fig:fig3.eps}
Numerical results have been shown for case (vi). As before the horizontal 
points showing uniform distribution (+) (for $\gamma=1$) have been drawn for reference.
$N$ is 5,000. The game has been played 20,000 times i.e. $D$ = 20,000. 
The uppermost panel shows the cases where agents depend on most recent
history only. The discount factor $\delta$ is set to $10^{-5}$. 
The middle panel shows the cases where agents assigns equal weightage on each and every game played in
history. The discount factor $\delta$ is set to 1. 
The last panel shows the cases where agents depend on the begining of the
history only. The discount factor $\delta$ is set to 1.1. In all cases the distributions of the fixed 
points have been shown for $\gamma=0.9$ ($\times$) and $0.7$ ($\ *$ ).
}
\end{center}
\end{figure}

\medskip

\section {variation in the ratio of the queue-lengths}

\noindent We consider the variation in the ratio of the queue-lengths and we try to find out numerically
the nature of the distribution of the ratio of the queue-lengths. This far we have 
found out the distribution of $Q_{Ai}$ and $Q_{Bi}$ only. Now we are concerned about the distribtion of
$L_A=(Q_{Ai}/Q_{Aj})$ and $L_B=(Q_{Bi}/Q_{Bj})$ where $j=i+1$. Clearly the distribution of 
the ratio of the fixed points acts as a proxy for the distribution of the ratio of the queue-lengths.
\medskip

{\bf Case(i)} {\it Agents follow the crowd (P\'olya model)} : Clearly  $P_{Ai}\sim Uniform(0,1)$
        
So  $L_A=(Q_{Ai}/Q_{Aj})$ ( where $j=i+1$ ) is the ratio of two uniformly distributed random variables.
 It can be shown that  $P(L_A)\sim(L_A)^{-2}$ for ($L_A>1$).

Numerical variations shown in the inset of fig. 4.
\medskip

{\bf Case(ii)} {\it Existence of a general preference} : It is seen from the above 
simulations that as the preference
parameter $\alpha$ goes up the density of fixed points favoring restaurant A increases. But
 no analytical study could be made for describing the distribution. The variations in the ratio 
of queue-lengths for different
values of the parameter are shown above. 
See fig. 4.
It should be noted that in the log-log plot the distribution appears to be
a straight-line for each parameter value, indicating the presence of a power-law.

\begin{figure}
\begin{center}
\noindent \includegraphics[clip,width= 5cm,angle = 270]
{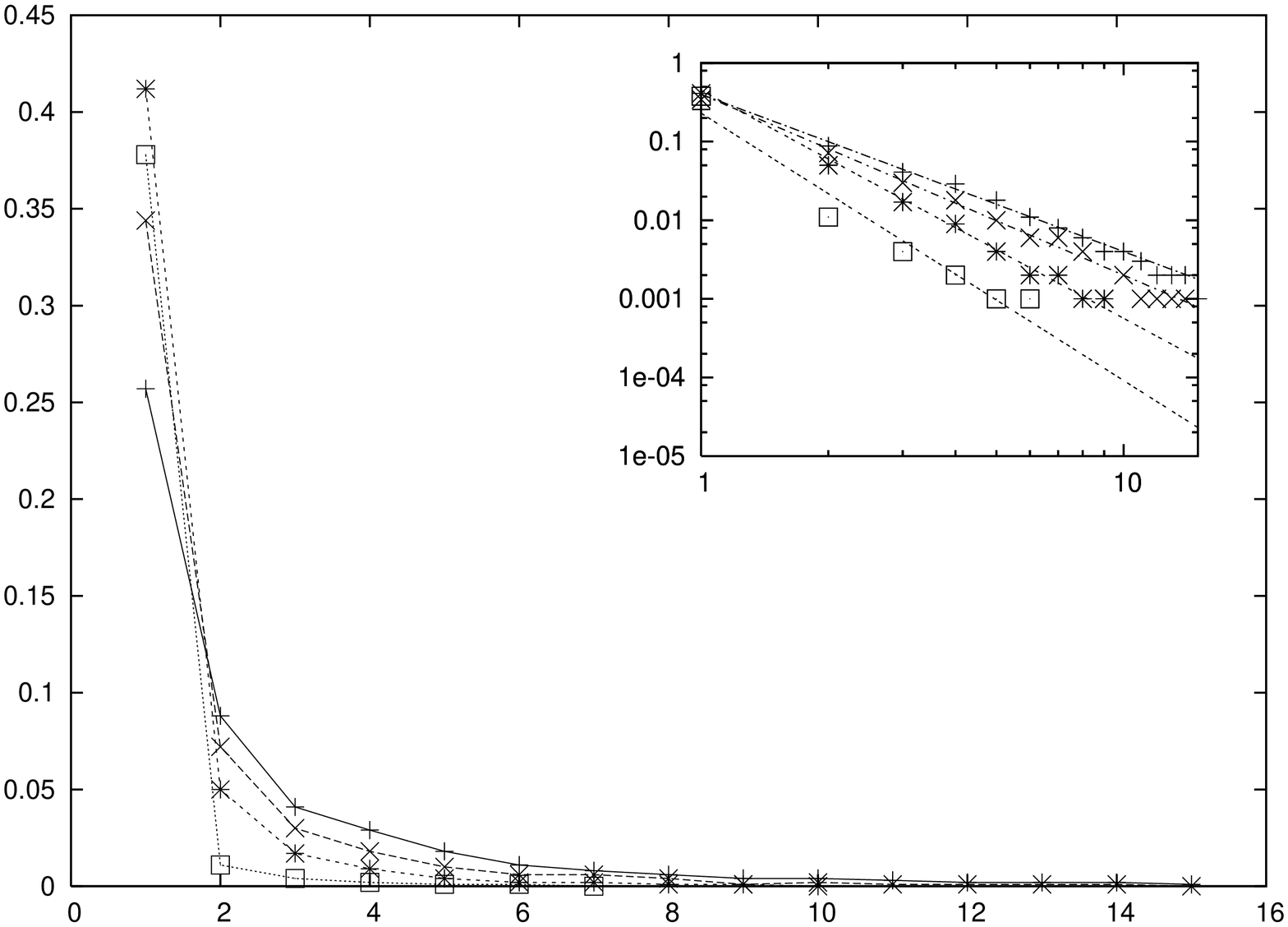}
\caption\protect{\label{fig:fig4} $N$ = 5,000 and $D$ = 10,000.
 The ratio of the queue-lengths has been plotted for $\alpha$=0
, 0.3, 0.6 and 0.9 with `+', `$\times$', `$\ *$ ' and `$\square$' repectively. 
It is to be noted that for $\alpha=0$ the 
problem reduces to P\'olya model. $\it Inset$ : The same figure is plotted on a log-log plot.
Straight-lines indicate the presence of power-laws. The uppermost straight-line 
has a slope = $-2$ (P\'olya model).
}
\end{center}
\end{figure}

\medskip

{\bf Case(iii)} {\it Existence of an arbitrary preference} : It is seen that with increasing
value of $\alpha$ the distribution of the queue-length converges to a $\delta$-function. 
The variations in the ratio of 
queue-lengths for different
values of $\alpha$ is shown below. 
See fig. 5. With $\alpha$ = 0,
the distribution follows a power-law with an exponent value -2, as has been discussed in case (i).
 With increasing absolute value of $\alpha$, it is seen in the result that the distribution becomes 
more peaked.

\begin{figure}
\begin{center}
\noindent \includegraphics[clip,width= 5cm,angle = 270]
{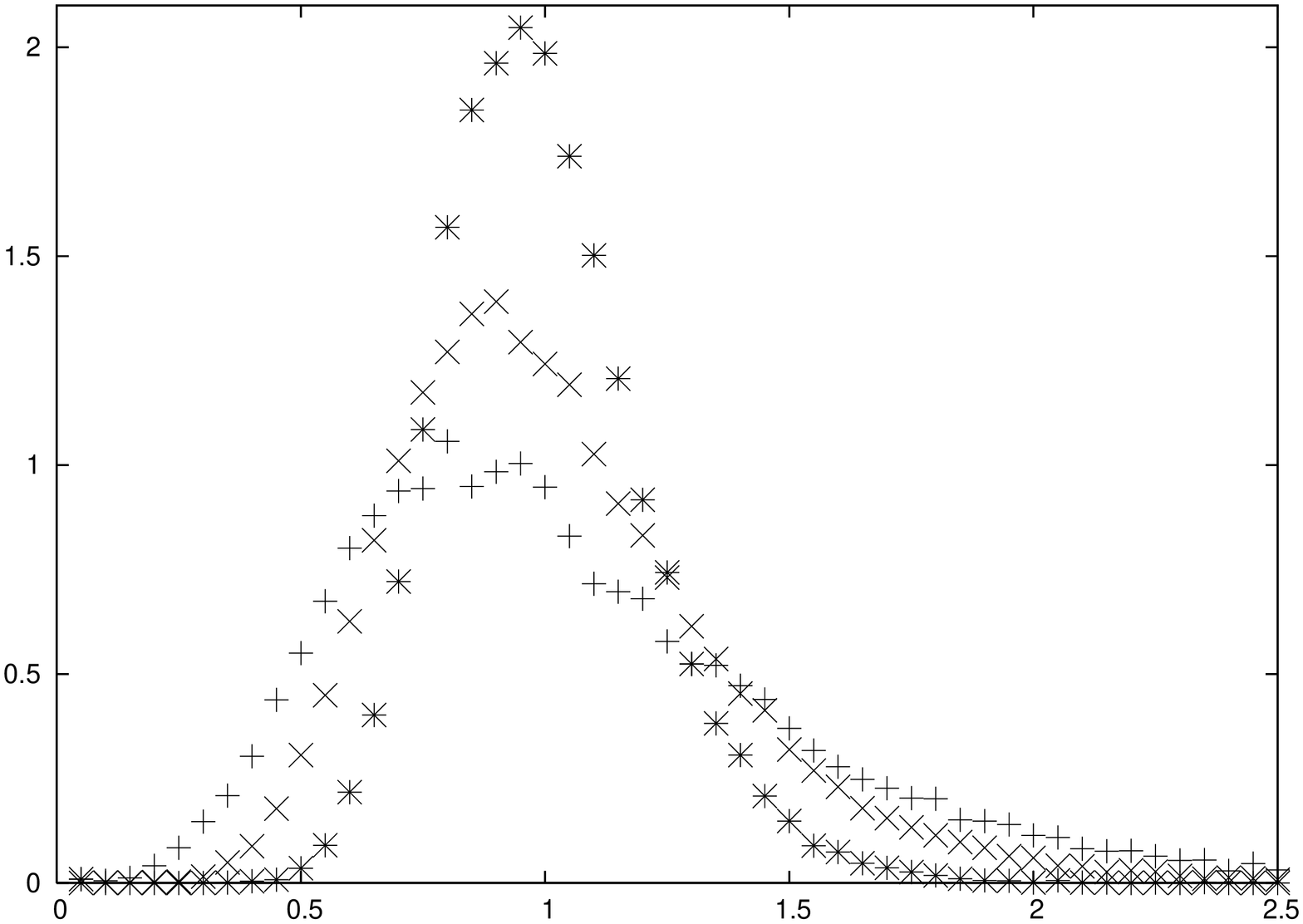}
\caption\protect{\label{fig:fig5}$N$ = 5,000 and $D$ = 20,000.
 The distribution of the fixed points are shown here for different absolute
values of the parameter, $|\alpha|$ = 10, 20, 50 with `+', `$\times$' and `$\ *$ '  respectively.
}
\end{center}
\end{figure}

\medskip

{\bf Case(iv)} {\it Dependence on history} : In this case also the distribution converges to
a $\delta$-function with decreasing values of $\gamma$ given $\delta$.
The variations in the ratio of queue-lengths for different
values of the parameters are shown below.
(See fig. 6). 
Just as before, with $\gamma$ = 1, the
 distribution follows a power-law with an exponent value -2, as has been discussed in case (i).
But for given $\delta$, it becomes more peaked with decreasing values of $\gamma$. 

\begin{figure}
\begin{center}
\noindent \includegraphics[clip,width= 12cm,angle = 270]
{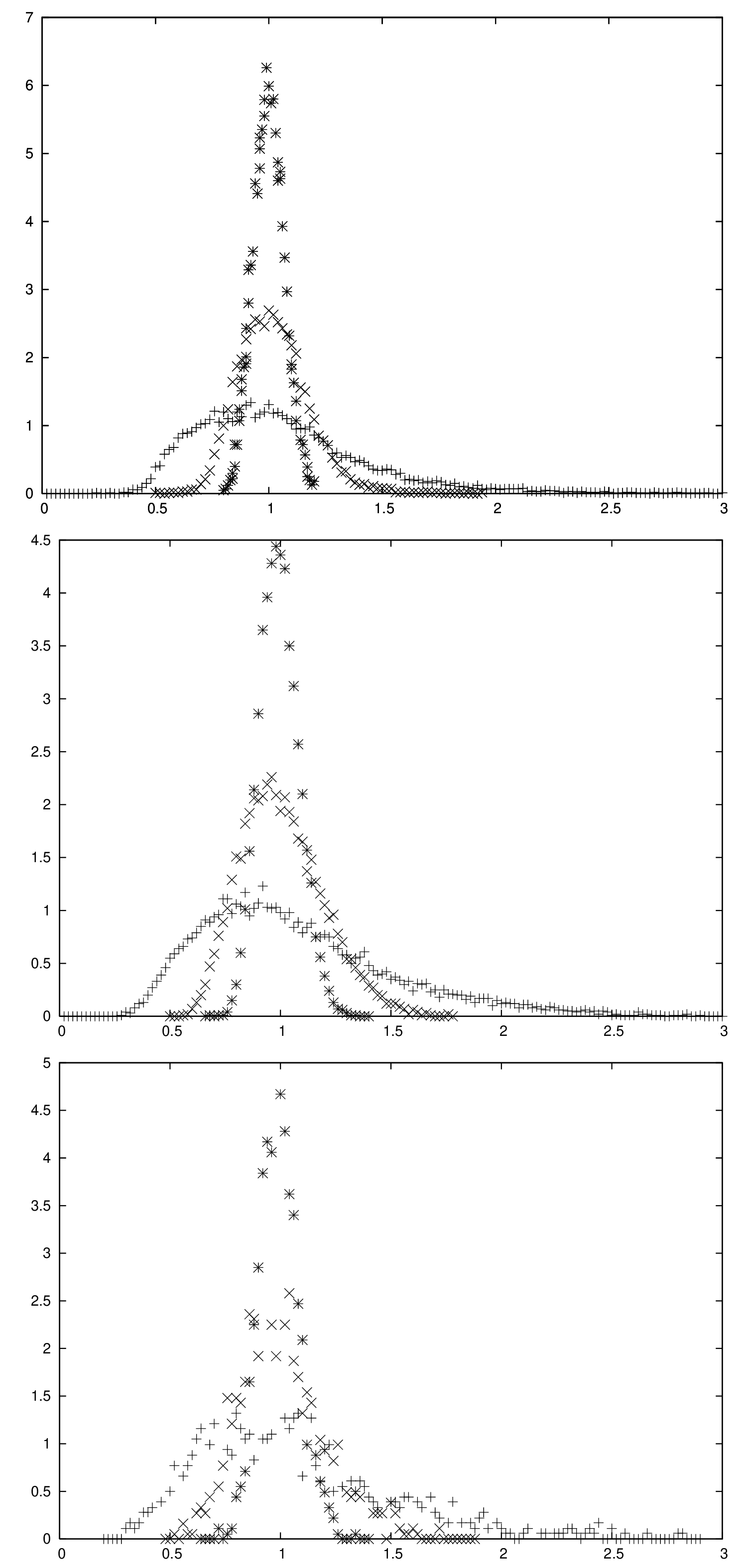}
\caption\protect{\label{fig:fig6}
In the above figures $N$ = 5,000 and $D$ = 15,000.
The distributions of the fixed points has been shown for $\gamma$= 0.9, 0.8 and 0.7 
with `+', '$\times$' and `$\ *$ ' respectively.
In the uppermost panel, agents remember the most recent history only ( $\delta = 10^{-5}$ ).
In the middle panel, agents assign an equal weightage to all results of the games played in past ( $\delta$ = 1 )
 and in the last panel, agents remember the begining of the history only ( $\delta$ = 1.1 ).
}
\end{center}
\end{figure}

\section {summary}
\bigskip

\noindent In this paper, we have investigated the distribution of the queue-lengths in P\'olya's Urn model and 
in several variations of it. Only
in the case of P\'olya model with a power greater than one (see eqn.1), the distribution collapses
to a $\delta$-function. Otherwise, there exists a non-trivial distribution over the range [0,1] no matter
what the strategy is. It is interesting to note that complete switching-over from one restaurant
to another is so rare in this model.
In the case where agents are influenced by history (see eqn.3) it is seen that the parameter $\delta$ does not play
much of an important role. The shape of the distribution is determined by the values of $\gamma$ 
which is actually the weight given to combined history
relative to the present scenario.

The existence of the power law in the ratio of the queue-lengths is also seen from these models.
The basic P\'olya model is in fact able to produce such a power law. We see the same for some variations
of it as well.
 It is to be noted that the value of the power derived in the case of P\'olya model (see case (i) above)
closely resembles the value observed in Smethurst et al [6] (see Freckleton et al [7]). 



\bigskip
\noindent
{\bf REFERENCE:}
\medskip

\noindent[1] A.V. Banerjee, A Simple Model of Herd Behaviour, Quart. J. Econ., {\bf 107} (1992) 797-817
\medskip

\noindent[2] A. Orl\'ean, Bayesian Interaction and Collective Dynamics of Opinion:Herd Behaviour and Mimetic Contagion, 

\noindent J. Econ. Behav. Org., {\bf 28} (1995) 257-274
\medskip

\noindent[3] D. Sornette, {\it Why Stock Markets Crash?},(Princeton University Press, New Jersey, 2003)
\medskip

\noindent[4] G. Weisbuch, Social Opinion Dynamics, in {\it Econophysics and Sociophysics:Trends and Perspectives},

\noindent Eds.: B.K. Chakrabarti, A. Chakraborti, A. Chatterjee (Wiley-VCH, Weinheim, 2006)
\medskip

\noindent[5] J.N. Lloyd and S. Kotz, {\it Urn Models and Their 
Applications:An Approach to Discrete Probability Theory}, 

\noindent (Wiley, New York, 1977)
\medskip

\noindent[6] D.P. Smethurst and H.C. Williams, Power Laws:Are Hospital Waiting Lists Self-regulating?, Nature, 
{\bf 410} (2001)

\noindent652-653
\medskip

\noindent[7] R.P. Freckleton and W.J. Sutherland, Do Power Laws Imply Self-regulation?, Nature,
{\bf 413} (2001) 382

\medskip

\noindent[8] F. Chung, S. Handjani and D. Jungreis, Generalizations of Polya's Urn Problem, Annals of 
Combinatorics, {\bf 7} (2003) 141-153

\medskip

\noindent[9] R. Pemantle, A Survey of Random Processes with reinforcement, Prob. Surveys,
{\bf 4} (2001) 1-79 

\end{document}